\documentclass[11pt,onecolumn,amssymb,nofootinbib]{revtex4}
\usepackage{amsmath}
\usepackage{amssymb}
\usepackage{graphicx}
\newcommand\beq{\begin{equation}}
\newcommand\eeq{\end{equation}}
\newcommand\beqa{\begin{eqnarray}}
\newcommand\eeqa{\end{eqnarray}}

\begin{document}
\title{Minimum measurement time: lower bound on the frequency cutoff for collapse models}

\author{Stephen L. Adler}
\email{adler@ias.edu} \affiliation{Institute for Advanced Study,
1 Einstein Drive, Princeton, NJ 08540, USA.}

\author{Angelo Bassi}
\email{abassi@units.it} \affiliation{Department of Physics, University of Trieste, Strada Costiera 11, 34151 Trieste, Italy\\ Istituto Nazionale di Fisica Nucleare, Trieste Section, Via Valerio 2, 34127 Trieste, Italy}

\author{Luca Ferialdi}
\email{lferialdi@units.it} \affiliation{Department of Physics, University of Trieste, Strada Costiera 11, 34151 Trieste, Italy\\ Istituto Nazionale di Fisica Nucleare, Trieste Section, Via Valerio 2, 34127 Trieste, Italy}

\begin{abstract}
The CSL model predicts a progressive breakdown of the quantum superposition principle, with a noise randomly driving the state of the system towards a localized one, thus accounting for the emergence of a classical world within a quantum framework. In the original model the noise is supposed to be white, but since white noises do not exist in nature, it becomes relevant to identify some of its spectral properties. Experimental data set an upper bound on its frequencies, while in this paper we bound it from below. We do so in two ways: by considering a `minimal' measurement setup, requiring that the collapse is completed within the measurement time; and in a measurement modeling-independent way, by requiring that the fluctuations average to zero before the measurement time.
\end{abstract}

\maketitle

\section{Introduction}

Collapse models are a phenomenological solution to the quantum measurement problem, where  the Schr\"odinger dynamics is modified by  adding non-linear stochastic terms, which trigger the collapse of the wave function~\cite{rep1,rep2}. In the widely popular mass-proportional CSL model~\cite{csl}, two parameters control the collapse: a collapse rate $\lambda$ and the noise space correlator $r_C$. Their numerical values are chosen in such a way that the collapse  is negligible for microscopic objects, thus recovering the standard quantum predictions, and become increasingly stronger for larger objects. While there is a generic consensus on the potential value of $r_C$ ($\sim 10^{-7}$m), the value of $\lambda$ is rather open~\cite{Aic91,Adl07}. If one requires the collapse to be effective at the level of latent image formation~\cite{Adl07}, then $\lambda \sim 10^{-8}$ s$^{-1}$, with an uncertainty of about 2 orders of magnitude. This is the value we consider in this article.

The original CSL noise is  white in time, because it is easier to study. However,  since white noises do not exist in nature,   it becomes relevant to consider the possibility of a non-white CSL noise, with a frequency cut off $\omega_M$. This has been done in~\cite{BasGhi02,AdlBas07,AdlBas08,BasDon14,CFB18} (for the extension to other models see e.g.~\cite{BasFer09a,BasFer09b,FerBas12a,FerBas12b}). The cutoff $\omega_M$ is a new phenomenological parameter; in this paper we aim at bounding it from below, by the requirement that the collapse be effective in measurement situations. We do so by following two distinct paths. On the one hand we consider the measurement process performed by a minimal experimental setup, consisting of a detector, an amplifier, a recording device, and a Lithium-ion battery. On the other hand, we bound the cutoff values by considerations on  the noise fluctuations, which should average to zero over a measurement process.

\section{Collapse rate for a non-white correlation function}
According to the non-white CSL model, the off-diagonal elements of the density matrix  $\langle r_1|\rho(t)|r_2\rangle $ decay exponentially, with exponent given by Eq.~(17) of~\cite{AdlBas08}:
\beq\label{Gamma}
\Gamma(t)= \lambda\frac{r_C^3}{\pi^{3/2}m^2_0}\int d^3x\int d^3y\int_0^t ds\, F(x-y,s)\Delta m(x)\Delta m(y)\,,
\eeq
where $m_0$ is the nucleon mass, $\Delta m(x)=m_1(x)-m_2(x)$ is the difference of the eigenvalues of the mass density operator for the states $|r_{1,2}\rangle$, and $F(x-y,t)$ is the time integral of the noise correlation function $D(x-y,t-s)$:
\beq
F(x-y,t)=\int_0^tds\,D(x-y,t-s)\,.
\eeq
We assume that $D(x-y,t-s)$ factorizes into spatial and temporal components, and that the noise has a high-frequency cutoff:
\beq
D(x-y,t-s)=\frac{\pi^{3/2}}{r_C^3}e^{-\frac{(x-y)^2}{4r_c^2}}\cdot\delta_{\gamma(\omega)}(t-s)\,,
\eeq
where the noise spatial correlator takes a Gaussian form, and $\delta_{\gamma(\omega)}(t-s)$ is the time correlator later defined.
Accordingly, Eq.~\eqref{Gamma} can be rewritten as follows
\beq\label{gammafull}
\Gamma(t)=\frac{\lambda}{m^2_0}\Lambda(t) \int d^3x\int d^3y\,e^{-\frac{(x-y)^2}{4r_c^2}}\Delta m(x)\Delta m(y)\,,
\eeq
where we have defined
\beq
\Lambda(t)=\int_0^tds\int_0^sdu\,\delta_{\gamma(\omega)}(s-u)\,.
\eeq
The temporal correlation $\delta_{\gamma(\omega)}$ is defined in terms of the cutoff function $\gamma(\omega)$ as follows:
\beq
\delta_{\gamma(\omega)}(t-s)=\frac{1}{\pi}\int_0^{\infty}d\omega\,\gamma(\omega)\cos[\omega(t-s)] .
\eeq
Note that when $\gamma(\omega)$  is constant (that is, $\gamma(\omega)=1$, no cutoff), one recovers the white noise case (Dirac delta correlation); in this case, $\Lambda(t)=t/2$.
There exists a number of cutoff functions that allow to evaluate $\Lambda(t)$ analytically, but the respective values of $\Lambda(t)$ differ only little (see Appendix A). For simplicity in the following we will use the Lorentzian cutoff function, because the associated $\Lambda(t)$ has a simpler analytic expression:
\beq\label{lambdalor}
\gamma(\omega)=\frac{\omega_M^2}{\omega^2+\omega_M^2}\qquad\rightarrow\qquad\Lambda(t)=\frac{ e^{-\omega_M t} + \omega_M t-1}{2 \omega_M}\,.
\eeq

\begin{figure}[h]
     \includegraphics[width=0.5\textwidth]{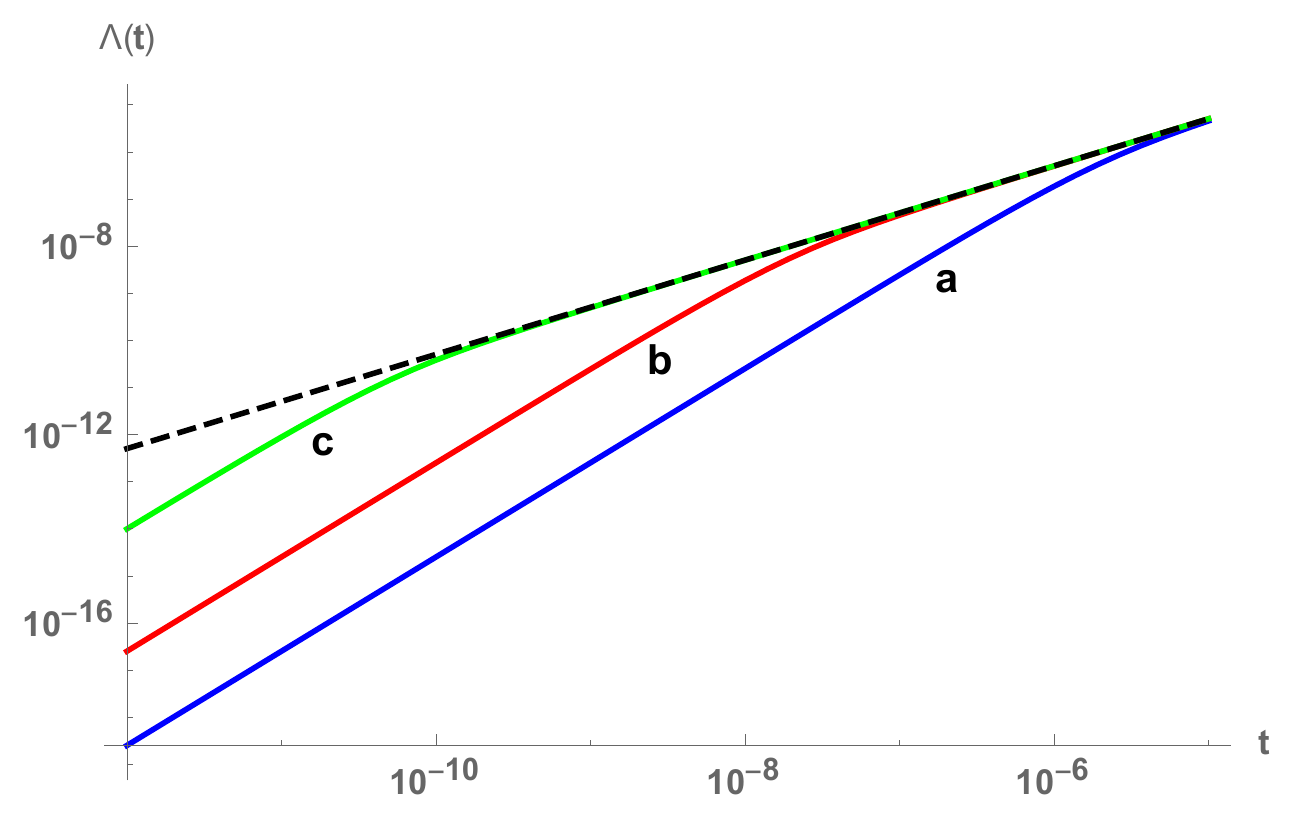} 
   \caption{$\Lambda(t)$ as a function of time, with Lorentzian cutoff, for different values of frequency cutoff. Solid lines denote: (a/blue) $\omega_M=10^6$~s$^{-1}$, (b/red) $\omega_M=10^8$~s$^{-1}$, (c/green) $\omega_M=4\times10^{10}$~s$^{-1}$. The lower  $\omega_M$, the longer it takes to reach the white noise rate (dashed line).
   }
    \label{loromega}
\end{figure}

Figure~\ref{loromega} compares $\Lambda(t)$ for different values of frequency cutoff, with the white noise case. One immediately sees that the lower the frequency cutoff, the smaller the collapse effect and the longer it takes to reach the same reduction rate as for the white noise case. The plot further shows that $\Lambda(t)$ has a ``sub-white'' behavior for $t\lesssim\omega_M^{-1}$, while the rate becomes effectively white for $t\gtrsim\omega_M^{-1}$.

If the frequency cutoff $\omega_M$ is too low, the collapse becomes ineffective. Therefore a lower bound exists for $\omega_M$, below which measurements do not return definite outcomes within their measurement time. The inverse cutoff thus identifies the minimum measurement time. Note that the measurement time, i.e. the time after which a single outcome should be produced, should not be identified with the resolution time, which quantifies the ability of the detector to distinguish two signals, potentially very close to each other. The second can be much smaller than the first, meaning that two distinct signals remain superimposed till the measurement is completed.

\begin{figure}[h]
 \includegraphics[width=0.5\textwidth]{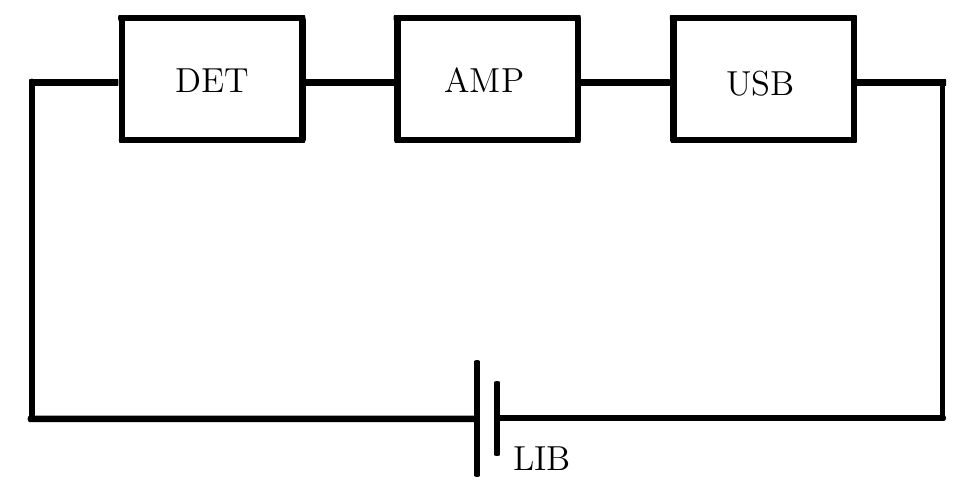}
    \caption{Minimal experimental setup considered. DET = detector
    ; AMP = amplifier; USB = flash drive; LIB = Lithium-ion battery.}
    \label{exp}
\end{figure}

We consider the extreme case of a  photon in a superposition state of hitting or not hitting the detector, which is then measured by the minimal experimental setup depicted in Fig.~\ref{exp}. This consists of a single photon detector (such as e.g. photomultiplier tubes, avalanche photodiodes, or superconducting nanowires \cite{Em2522,mcp}), an amplifier~\cite{ADN2880,Nol18}, and a flash drive that records the signal \cite{web}. The setup is powered by a Lithium-ion battery (LIB).

A setup of this kind represents a minimal idealized scenario, as it involves the displacement of the least number of nucleons. As such, the analysis of this setup will provide  the strongest bounds on the cutoff frequency. In performing this analysis, our essential assumption is that a measurement is completed by the time a permanent record of the event is made, that can be read out at a later time.

Each part of the setup has a different time scale, namely the detection time $t_D$, the amplification time $t_A$, and the recording time $t_R$. We thus identify the measurement time $t_M$ with the sum of these time scales. In our analysis, we consider a detection time of $10^{-8}$ s, which can be easily reached with commercial components, although shorter timescales are achievable with more advanced setups~\cite{footnote1}.

We first try the more stringent requirement that the measurement is completed when the detector produces an output ($t_M=t_D$). We can realistically assume that the current pulse produced by the detector has width $t_P$ one order of magnitude smaller than the detection time.  Since the collapse model we are considering is mass proportional, we are interested in computing the number of ions displaced in the LIB in the measurement process (the electron contribution is negligible). In a typical LIB, Li$^+$ ions diffuse with velocity $v\simeq0.28\times10^{-4}$ cm/s~\cite{vel} in a 1 molar solution of LiPF$_6$ electrolyte, whose thickness is about $h\simeq10^{-2}$ cm~\cite{electrolyte}. 
The number of ions in the solution displaced inside the battery when a current $I$ is generated, can be calculated starting from the formula $I=\rho v A$, where $\rho$ is the ion density in the electrolyte, and $A$ is the electrode surface between which the flux occurs. Since $\rho= N/(A h)$, one finds that the number $N$ of ions displaced by $I$ is $N=Ih/v$.
For a current pulse of 2 mA, corresponding to the typical output current of a single photon detector (100 mV @ 50 $\Omega$)~\cite{footnote1}, one gets $N=4.46\times10^{18}$ ions.
If we consider a conservative detection time $t_D=10^{-8}$ s, we have that $t_P=10^{-9}$ s, and each ion travels an average distance $\Delta=0.28\times10^{-13}$ cm.

In order to evaluate the collapse rate, we assume that the i-th ion in the electrolyte, centered around position $r^1_i$ if no detection occurs, travels a distance $\Delta_i$ to the new position $r^2_i=r^1_i+\Delta_i$ if detection occurs and the current is generated to power the circuit. For displacements much shorter than $r_C$, one can rewrite Eq.~\eqref{gammafull} as follows
\beq\label{gammatrans}
\Gamma(t)=\frac{\lambda r_C^3m^2}{\pi^{3/2}m^2_0}\Lambda(t)\sum_{i,j=1}^N\int d^3k\,(\Delta_i\cdot k) (\Delta_j \cdot k)e^{-r_c^2k^2}e^{-ik\cdot(r_i^1-r_j^1)}\,,
\eeq
where we have assumed a discrete mass distribution $m_{1,2}(x)=m\sum_i \delta(x-r^{1,2}_i)$.
Since individual ion motions are uncorrelated, we estimate Eq. \eqref{gammatrans} by keeping the positive definite  $i=j$ diagonal terms, and ignoring the oscillatory
$i\neq j$ off-diagonal terms.  This gives, with $m=nm_0$ and $\sum_i 1=N$,
\beq\label{n2N}
\Gamma(t)=\lambda \,n^2\,N\,\frac{\Delta^2}{2r_C^2}\,\Lambda(t)\,,
\eeq
where  $n$ is the number of nucleons in each ion. We remark that the factor $\Delta^2/2r_C^2$ accounts for the fact that only part of the nuclear mass contributes to the collapse factor (see Appendix B for a detailed explanation). Recalling that $N=I h/v$ and $\Delta=vt$, and by taking into account the fact that both Li$^+$ and PF$_6^-$ ions are displaced by the electric potential,  one gets
\beq\label{gammaav}
\Gamma(t)=\lambda \left(n_{\mathrm{Li}}^2+n_{\mathrm{PF_6}}^2\right)\frac{I\,h\,v}{2r_C^2}\,t^2\,\Lambda(t)\,,
\eeq
with $n_{\mathrm{Li}}=7$ and $n_{\mathrm{PF_6}}=145$.
If we assume that the reduction is complete when\footnote{ This choice, which is the one usually considered in the literature~\cite{Adl07}, corresponds to reduction occurring one term of the incoming superposition is $e$ times bigger than the other.} $\Gamma(t)\approx1$, recalling that for high frequencies $\Lambda(t)\sim t/2$ (white noise), we find that for Adler's value of $\lambda$, the collapse (or reduction) time is $t_C=8.16\times10^{-6}$ s. This is about three orders of magnitude too large to guarantee collapse at the detection stage.

Although for the considered detection time scales the collapse does not occur, we recall that detector signals are typically amplified before being shown on an oscilloscope or registered on a hard drive.
Motivated by the previous results, we then extend our analysis by requiring the collapse to be completed when one bit of information is recorded by a NAND flash memory.
We thus assume that the detector output current ($\sim 2$ mA) is amplified as much as needed by a NAND for write operations.
Amplifiers can provide the needed gain on the time scale $t_A=10^{-8}$ s (commercial components) or even faster (advanced devices)~\cite{footnote1}.
Commercial NANDs have a write latency of about $t_R=10^{-4}$ s~\cite{samsungNAND,intelNAND}, while latest generation devices perform one order of magnitude faster ($t_R=10^{-5}$ s), with a current consumption of about 13.8 mA at least~\cite{NANDcurr}. Multiple NANDs are incorporated in flash drives or SSDs, which require a current of about 500 mA for the write operation.\footnote{The currently fastest flash drive has a power consumption of about 2.7 W~\cite{SSD}. For a 5 V rail, this corresponds to a current of 540 mA. Since the device is USB2 compatible, we can safely assume the current not to be higher than the maximum for a USB2. The latest generation SSDs~\cite{samsungNAND} have a power consumption of 9W with a 12V rail, corresponding to a current of 750mA. In our estimates we use the more conservative value of 500mA.  A multi-gigabyte flash drive
has more than $10^{10}$ NANDs, each of which records one bit. A data transfer rate of $\sim 10^9$ bits per second is a factor of $10^4$ higher than the inverse latency
time for a NAND to write a bit into a permanent record; thus one kilobyte of data can be transferred during the write latency time.   The current ratio (500 mA/13.8 mA) $\sim$ 36 is much smaller than $10^4$ because long strings of NANDs are connected
in series to the 5V rail \cite{wiki}.}

 Since flash drives record on a time scale much longer than the detection and amplification processes, we redo our analysis by identifying the measurement $t_M$ time with the record time $t_R$. We will thus consider two values for $t_M$ ($10^{-5}$ s and $10^{-4}$ s), and two values for the current needed for recording ($13.8$ mA and $500$ mA).
The total number of ions displaced inside the battery when the current is generated, amounts to $N=1.11\times 10^{21}$ for a $500$ mA current, and $N=3.08\times10^{19}$ for a $13.8$ mA current.

In the white noise case, Eq.~\eqref{gammaav} tells that reduction occurs at $t_C= 1.30\times10^{-6}$ s for a 500 mA current, and at $t_C=4.29\times10^{-6}$ s for a 13.8 mA current\footnote{In providing these estimates, we have assumed that the PF$^-_6$ ions diffuse with the same velocity as Li$^+$ ions, because the PF$^-_6$ diffusion velocity is not available in the literature. However, one might estimate it by assuming momentum conservation, which gives a PF$^-_6$ velocity about 20 times smaller than Li$^+$. Since Eq.~\eqref{gammaav} depends linearly on the ion velocity, and considering that in the white noise case the collapse time goes with the cube root of $\Gamma(t)$, then taking the lower PF$^-_6$ velocity gives a correcting factor of about 2.7. This numerical factor does not significantly change our estimate.}. Both collapse times are smaller than the measurement time, implying that a classical record is achieved before the end of the measurement process.
The above reduction times were obtained using the value of $\lambda=10^{-8}$ s$^{-1}$ proposed by Adler~\cite{Adl07}; if one were using the value proposed by GRW~\cite{Aic91}, i.e. $\lambda=10^{-16}$ s$^{-1}$, the reduction times would become more than two orders of magnitude larger. Accordingly, such a small value of $\lambda$ would not guarantee a definite measurement outcome, not only for the minimal setup considered here, but also for more complicated setups\footnote{We remark that the value of $\lambda$ can in principle be decreased as long as the collapse time $t_C$ matches the  measurement time $t_M$. For the measurement time $t_M=10^{-5}$ s one can decrease $\lambda$ by about a factor $2.2\times10^{-3}$ for a 500mA current, and by a factor $7.9\times10^{-2}$ for 13.8mA. Similarly, if one considers $t_M=10^{-4}$ s, $\lambda$ can be made even smaller, respectively by $2.2\times10^{-6}$ (500mA) or $7.9\times10^{-6}$ (13.8mA).}.

\begin{figure}[h]
 \includegraphics[width=0.5\textwidth]{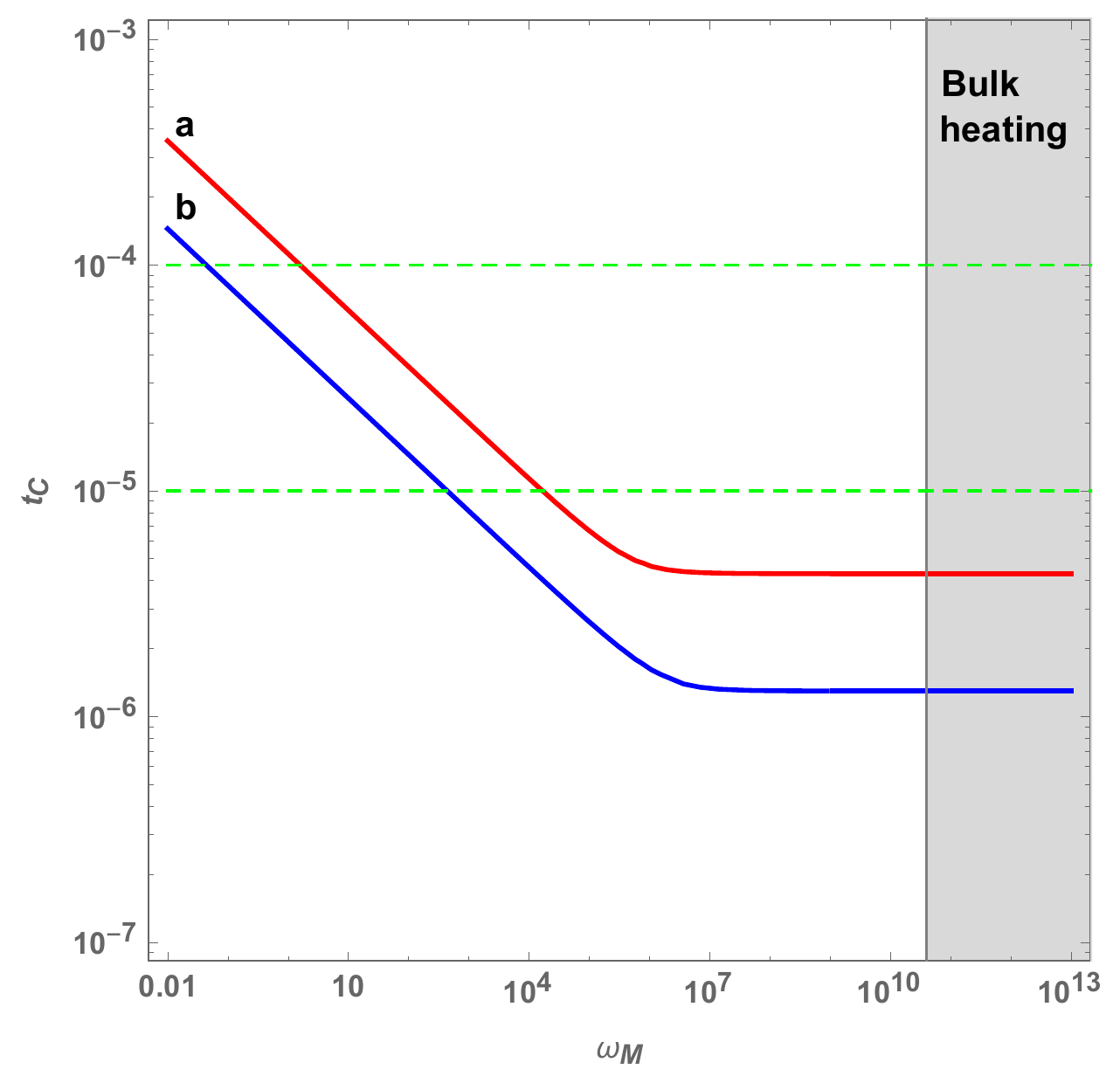}
    \caption{Collapse time $t_C$ (s) as a function of the frequency cutoff $\omega_M$ (s$^{-1}$), for the minimal measurement setup considered in the paper. The red (a) solid line refers to a setup where a current of 13.8 mA is used for the write operation. The blue (b) solid line refers to a current of 500 mA. Green dashed lines denote the two measurement time scales considered: $t_M=10^{-4}$ s and $t_M=10^{-5}$ s. The intersection between solid and dashed lines identifies the lower bound for the cutoff frequency $\omega_M$. The vertical gray line corresponds to the cutoff $\omega_M$ upper bound coming from bulk heating experiments~\cite{Adl07,bulk}, and the shaded area denotes the excluded parameter range.}
    \label{cp200}
\end{figure}

Bounds on the frequency cutoff are obtained simply by using Eq.~\eqref{lambdalor} for $\Lambda(t)$ in Eq.~\eqref{gammaav}.
Figure~\ref{cp200} shows the values that $t_C$ and $\omega_M$ can take in order to guarantee that the superposition of the incoming photon is suppressed ($\Gamma(t)\approx 1$). Reduction occurs within the measurement time only if the solid line crosses below the dashed line corresponding to the considered measurement time (i.e. $t_C< t_M$). The intersection between solid and dashed lines occurs when $\omega_M$ is large enough so that the collapse time $t_C$ (solid lines) becomes equal to the measurement time $t_M$ (dashed lines), thus identifying the lower bound for the cutoff frequency $\omega_M$.
We thus see that for a measurement time $t_M=10^{-4}$~s, a cutoff $\omega_M\gtrsim1$~s$^{-1}$ guarantees that the superposition is reduced within the measurement time when a $13.8$ mA current is involved (a: red solid line), while a cutoff $\omega_M\gtrsim5\times10^{-2}$~s$^{-1}$ is sufficient for a current of $500$ mA (b: blue solid line). As expected, a shorter measurement time $t_M=10^{-5}$~s sets stronger bounds, namely $\omega_M\gtrsim10^4$~s$^{-1}$ (for $13.8$ mA) and $\omega_M\gtrsim5\times10^2$~s$^{-1}$ (for $500$ mA).

For the sake of completeness, we have also investigated whether heating due to the current flow might contribute significantly to the collapse rate. In Appendix C we show that this is not the case since the phonon displacement due to heating is too small.

In order to overcome the difficulties in providing a precise estimate of the number of nucleons displaced by the measurement process, in the next section we consider a device-independent approach.

\section{Device-independent measurement time}
To find a device-independent measure of the collapse time, we consider the time it takes for the noise fluctuations to average to zero. This implies that the noise has fluctuated enough to lead the state vector to a (almost) definite eigenstate of the collapse operator\footnote{According to collapse models, when the state vector is initially in the superposition of different eigenstates of the collapse operator (in our case, the mass density operator), the noise makes the state fluctuate in time, driving it away from the superposition, towards  a localised state. Therefore, when the noise has fluctuated enough, a collapse occurs.}, thus producing a different outcome for different individual noise histories.  Let $\xi(t)$ be the real-valued noise as a function of the time $t$, normalized so that
\begin{equation}\label{noiseave}
\delta_{\gamma(\omega)}(t-s)=\delta_{\gamma(\omega)}(s-t)= \mathbb{E}\left[\xi(s) \xi(t)\right]~~~.
\end{equation}
 We consider two measures of the noise fluctuations defined by the average of the noise correlation over the time interval:
\beqa
\widetilde{I}(t)&\equiv& \, \mathbb{E}\left[\xi(t) \bar{\xi}(t)\right]=\frac{1}{t}\int_0^tds\,\delta_{\gamma(\omega)}(t-s)\,,\\
\widetilde{J}(t)&\equiv &\mathbb{E}\left[\bar{\xi}(t)^2\right]= \frac{2}{t^2}\Lambda(t)\,,
\eeqa
where $\bar{\xi}(t)$ is the noise average over the time interval:
\beq
\bar{\xi}(t)\equiv\frac{1}{t}\int_0^tds\,\xi(s)\,.
\eeq
$\widetilde{I}(t)$ measures the fluctuations of the noise realisation over a time $t$ with respect to its average, while $\widetilde{J}(t)$ measures the fluctuations of the noise average itself.
One can easily check that in the limit $t\rightarrow\infty$ both $\widetilde{I}(t)$ and $\widetilde{J}(t)$ tend to zero (fluctuations average out), while in the limit $t\rightarrow0$, the two functions approach a constant, which depends on the cutoff function. It is convenient to normalize $\widetilde{I}(t)$ and $\widetilde{J}(t)$ by the respective initial values, obtaining for a Lorentzian cutoff
\beqa
I(t)&\equiv& \widetilde{I}(t)/\widetilde{I}(0)=\frac{ 1}{\omega_M t}\,\left(1-e^{-\omega_M t} \right)\,,\\
J(t)&\equiv& \widetilde{J}(t)/\widetilde{J}(0)=\frac{ 2}{\omega_M^2 t^2}\,\left(\omega_M t-1+e^{-\omega_M t} \right)\,.
\eeqa
$I(t)$ and $J(t)$ thus take values between zero and one. In order to bound $\omega_M$, we need to choose a threshold below which we consider $I(t)$ and $J(t)$ small enough (ideally it should be $I(t)=0$, $J(t)=0$). It is reasonable to consider such a threshold to be $0.1$, meaning that when $I(t)$ and $J(t)$ go below this value, the noise has already fluctuated enough to give different measurement outcomes. Though reasonable, this is an arbitrary value, and if one were to choose a lower threshold, stronger lower bounds on $\omega_M$ would be obtained.

\begin{figure}[h]
     \includegraphics[width=0.5\textwidth]{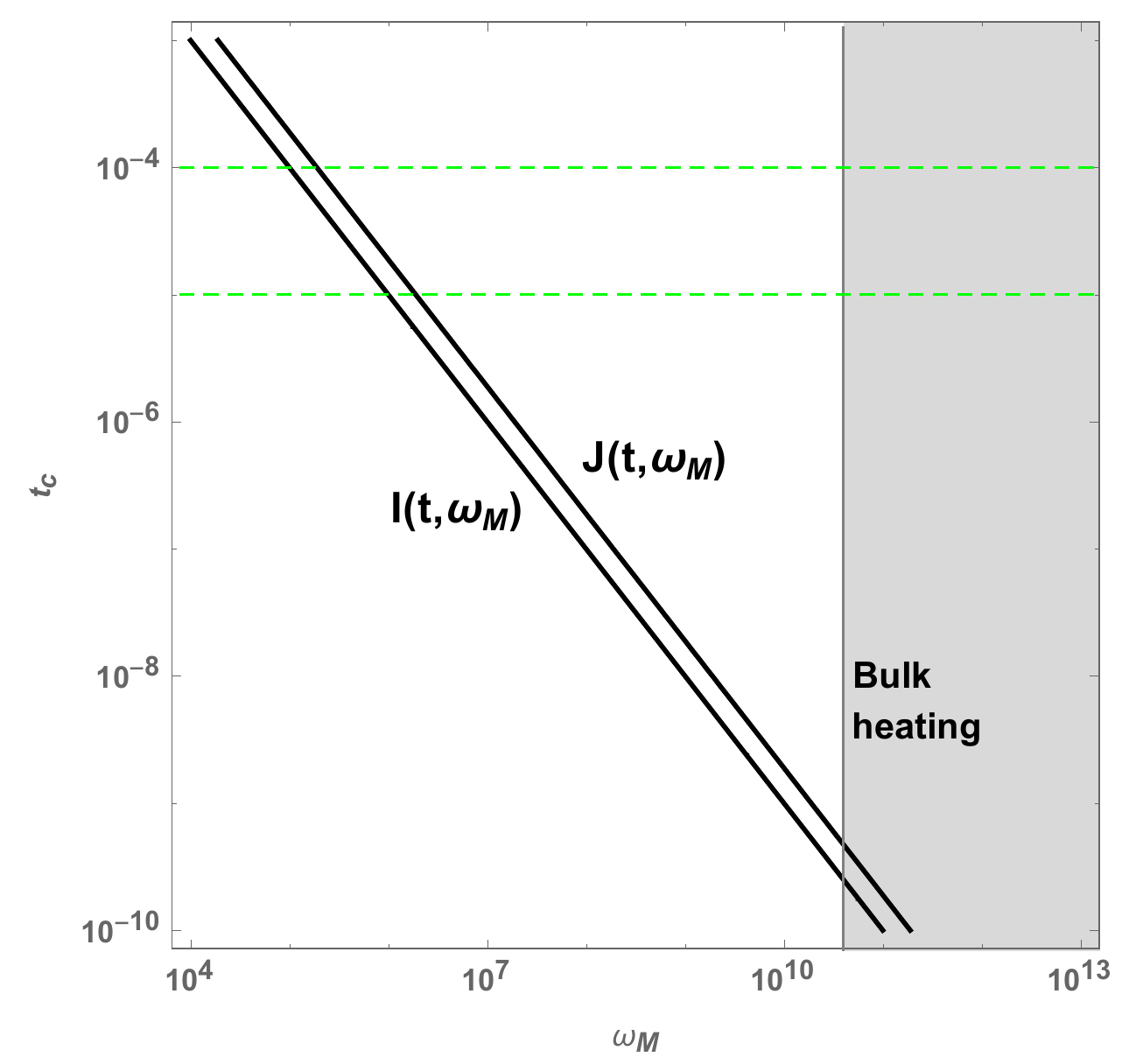} 
   \caption{Solid lines show when $I(t_C,\omega_M)$ and $J(t_C,\omega_M)$, defined in the main text, are equal to $0.1$.  Dashed green lines and gray are are as in Fig. 3. As for the previous figure, the intersection between solid and dashed lines identifies the lower bound for the cutoff frequency $\omega_M$.}
    \label{IJ}
\end{figure}

Figure~\ref{IJ} shows the values of $t$ and $\omega_M$ such that $I(t),J(t)=0.1$ (solid black lines). For each timescale considered, the lower bound on the frequency cutoff is given by the value that $\omega_M$ takes at the intersection between the solid lines with the respective dashed lines. One thus notices that, for the measurement time scale $t_M\leq 10^{-4}$ s, it is possible to have different collapse outcomes, provided that $\omega_M\gtrsim10^{5}$~s$^{-1}$. This bound is more than five orders of magnitude stronger than the one obtained in the previous section. Similarly, for a measurement time $t_M\leq 10^{-5}$ s, different measurement outcomes are guaranteed by a cutoff $\omega_M\gtrsim10^{6}$~s$^{-1}$, that is more than two orders of magnitude larger than the one obtained previously.  We suggest that the bounds of Sec. 2 are so much weaker 
than the present ones because the ion
drift velocity is so small that the ions move by much less than $r_C$ in the course of a measurement, leading to the $\Delta^2$ factor in Eq. \eqref{n2N}, and the consequent
$t^3$ behavior of $\Gamma(t)$ (for a white noise).  It is also possible that the analysis of the previous section underestimates the number of ions that move in the measurement process.

\section{Discussion and Conclusions}
In order to bound the frequency cutoff of the collapse noise of the CSL model, we have considered a minimal experimental setup detecting a single photon in a superposed state. We first investigated whether the collapse occurs by the time the detector produces an output, and we showed that this is not the case. We thus required the collapse to occur before the outcome is registered on a flash drive. Our analysis considered two different time scales: $t_M=10^{-4}$ s, that corresponds to the write time of commercial NANDs, and $t_M=10^{-5}$ s, achievable with the latest generation NANDs; and two different currents, 13.8 mA for a single NAND and 500 mA for a flash drive.  We found that for the measurement time scale $t_M=10^{-4}$ s, reduction is guaranteed for $\omega_M\gtrsim1$~s$^{-1}$ (13.8 mA) or $\omega_M\gtrsim5\times10^{-2}$~s$^{-1}$ (500 mA). The timescale $t_M=10^{-5}$ s gives slightly stronger bounds: $\omega_M\gtrsim10^{4}$~s$^{-1}$ (13.8 mA) or $\omega_M\gtrsim5\times10^2$~s$^{-1}$ (500 mA).

Since it is difficult to provide a more precise estimate of this kind, we also proposed a device-independent method to bound the frequency cutoff of the CSL noise. This method is based on the assumption that noise fluctuations should average to zero within the measurement time, indicating that the collapse is complete. By defining two fluctuation measures, and by requiring them to be small, we showed that reduction occurs within the measurement times $t_M= 10^{-4}$ s and $t_M= 10^{-5}$ s, provided that the frequency cutoff is respectively larger than $\omega_M\simeq10^{5}$~s$^{-1}$ and $\omega_M\simeq10^{6}$~s$^{-1}$.

These bounds might be further improved if nanoscale devices can be built that register a permanent record with write latency considerably shorter than $10^{-5}$s.  We pose
this as an experimental challenge for nanoscience research.




\section{Acknowledgements}
SLA wishes to thank the following people for very helpful information about fast electronic
measurements and NAND write time:  Jason W. Fleischer, William Happer, Kirk McDonald, Peter D. Meyers,
Margaret Murnane,  Mitch Newcomer, Nai Phuan Ong, Vinay H. Patel, Xiaohang Sun, Christopher G. Tully, Anthony Adler, and Jozef Mitros. SLA also wishes to acknowledge the Aspen Center for Physics, which is supported by National Science Foundation grant PHY-1607611, where part of this work was done.
AB and LF acknowledge financial support from the H2020 FET Project TEQ (grant n. 766900). AB also acknowledges financial support from INFN, FQXi, the COST Action QTSpace (CA15220), and hospitality from the Institute for Advanced Study, Princeton. AB and LF acknowledge Paolo Camerini, Silvio Modesti, Gianni Ramponi, Stefano Marsi and Giangiacomo Guerreschi and his colleagues at Intel for fruitful discussions. The authors wish to thank one of the anonymous referees for the many constructive comments, which helped in improving our work.

\section*{Appendix A: $\Lambda(t)$ for different cutoff functions}
\begin{figure}[h]
 \begin{minipage}[h]{0.4\textwidth}
     \includegraphics[width=\textwidth]{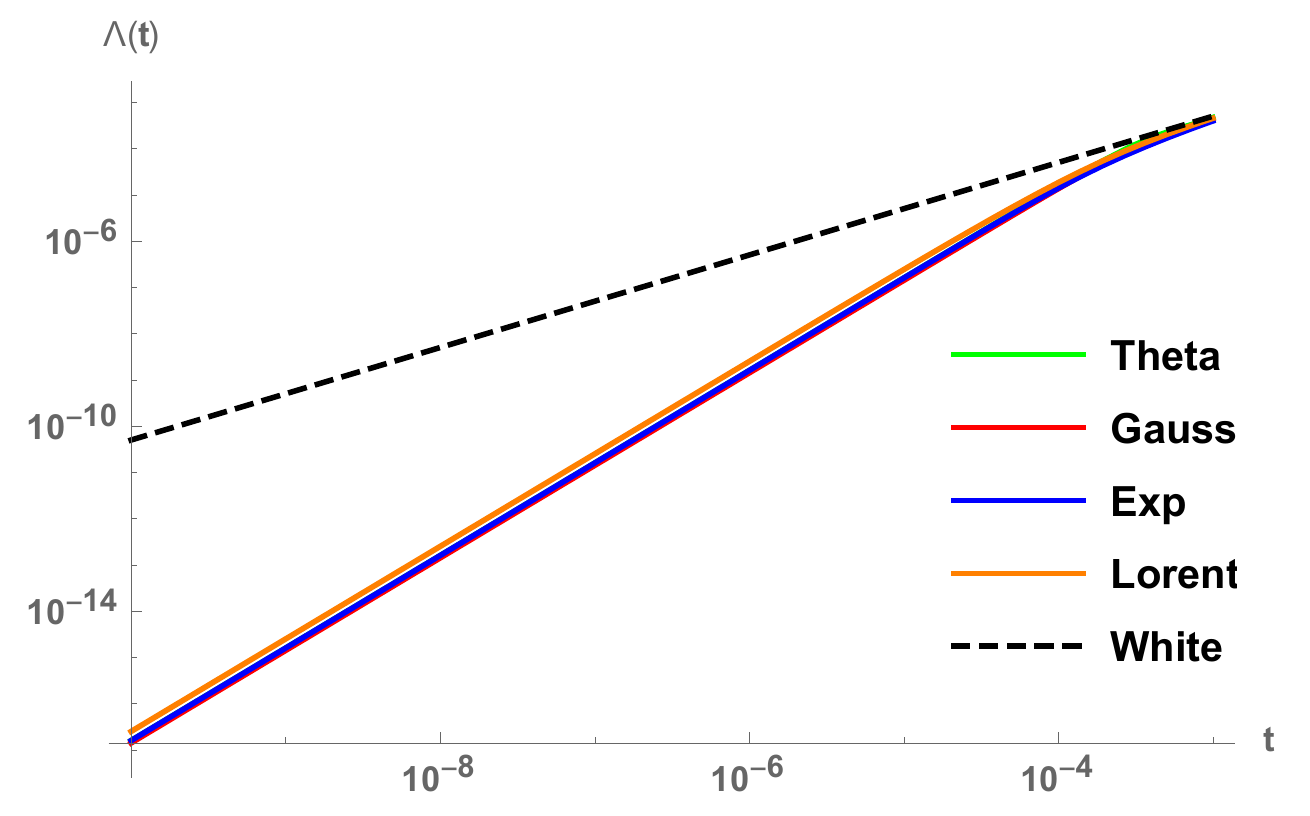} 
  \end{minipage}
  \hspace{1cm}
  \begin{minipage}[h]{0.4\textwidth}
     \includegraphics[width=\textwidth]{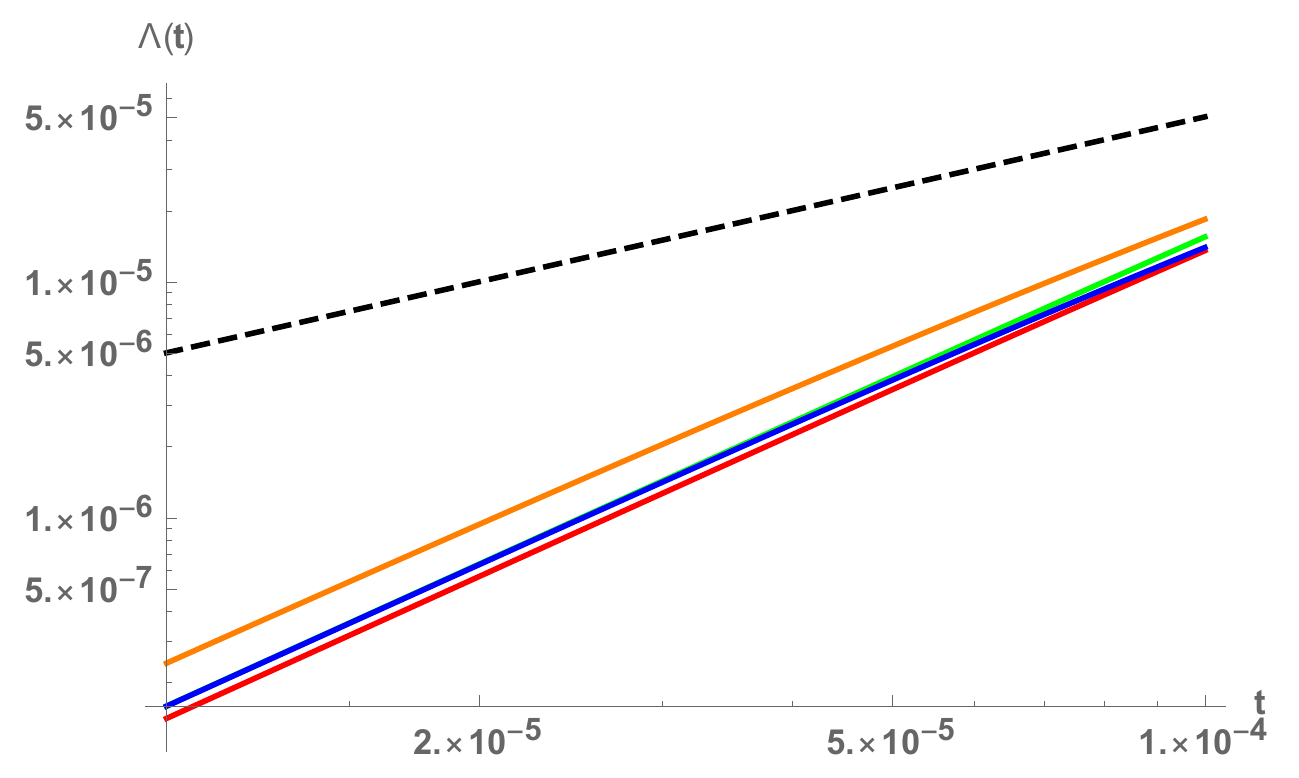}
  \end{minipage}
   \caption{Left figure shows $\Lambda(t)$ for different cutoff functions on the full range $t=10^{-10}-10^{-3}$ s for a (quite low) cutoff frequency $\omega_M=10^4$ s$^{-1}$. For higher cutoffs the lines are almost undistinguishable. The right plot is a zoom of the range $t=10^{-5}-10^{-4}$~s.}
    \label{cutoffs}
\end{figure}

There exist a number of cutoff functions that allow one to evaluate the time dependent part of the collapse rate analytically. These are some examples (with cutoff frequency $\omega_M$):
\beqa
\gamma(\omega)=\theta(\omega_M-\omega)\qquad&\rightarrow&\qquad\Lambda(t)=\frac{1}{\pi}\left[\frac{ \cos(\omega_M t)-1}{\omega_M} +  t \,\mathrm{Si}(\omega_M t)\right]\\
\gamma(\omega)=\exp\left(-\frac{\omega^2}{\omega_M^2}\right)\qquad&\rightarrow&\qquad\Lambda(t)=\frac{1}{\omega_M\sqrt{\pi}}\left[e^{-\frac{ \omega_M^2 t^2}{4}}-1\right] + \frac{t}{2} \mathrm{Erf}\left(\frac{\omega_M t}{2}\right)\\
\gamma(\omega)=\exp\left(-\frac{\omega}{\omega_M}\right)\qquad&\rightarrow&\qquad\Lambda(t)=\frac{1}{\pi}\left[t \,\mathrm{arctan}(\omega_M t) - \frac{\log(1 + \omega_M^2 t^2)}{2 \omega_M}\right] \\
\gamma(\omega)=\frac{\omega_M^2}{\omega^2+\omega_M^2}\qquad&\rightarrow&\qquad\Lambda(t)=\frac{ e^{-\omega_M t} + \omega_M t-1}{2 \omega_M}\\
\gamma(\omega)=1\qquad&\rightarrow&\qquad\Lambda(t)=\frac{t}{2}
\eeqa
The white noise limit is obtained for $\omega_M\rightarrow\infty$.
Figure~\ref{cutoffs} shows that the  behaviour of $\Lambda(t)$ at fixed cutoff frequency $\omega_M=10^4$ varies only little for the different cutoff functions. We remark that the higher $\omega_M$, the smaller the difference among the functions. In the main text we opted for the Lorentzian cutoff because the associated $\Lambda(t)$ has a simpler analytic expression.

\section*{Appendix B:} 
We show that Eq.~\eqref{n2N}, which was derived under the assumption of a point-like mass distribution for each nucleon, holds true also for an extended mass distribution. We start from Eq.~\eqref{gammafull} and we assume $m_{1,2}(x)$ to be two spherical mass distributions of radius $R$, whose centers are displaced by $\Delta$, say in the x direction. In this case Eq.~\eqref{gammatrans} thus reduces to
\beq\label{}
\Gamma(t)=\frac{\lambda r_C^3m^2}{\pi^{3/2}m^2_0}\Lambda(t)\,\Delta^2\int d^3k\, e^{-r_c^2k^2}k_x^2 |\mu(\boldsymbol{k})|^2\,,
\eeq
where $\mu(\boldsymbol{k})$ is the Fourier transform of the mass distribution. For a spherical mass the integral can be performed exactly and gives (see e.g. Eq.~(S.10) of~\cite{opto})
\begin{equation}
\int d^3k\, e^{-r_c^2k^2}k_x^2 |\mu(\boldsymbol{k})|^2=\frac{\pi^{3/2}}{r_C^5}\left[e^{-\frac{R^2}{r_C^2}}-1+\frac{R^2}{2r_C^2}\left(e^{-\frac{R^2}{r_C^2}}+1\right)\right]\frac{6r_C^6}{R^6}\,.
\end{equation}
By expanding the square brackets in the limit $R\ll r_C$ one finds $[\dots]\simeq\frac{R^6}{12r_C^6}$, thus giving
 \beq\label{n2Napp}
\Gamma(t)\simeq\lambda n^2\frac{\Delta^2}{2r_C^2}\Lambda(t)\,,
\eeq
where we have used $m=n\,m_0$. This is the same as Eq.~\eqref{n2N} for one nucleus.

The same is true for any spatially limited mass distribution, independently of its shape. For two mass distributions displaced by an amount $\boldsymbol{\Delta}$, Eq.~\eqref{gammafull} can be rewritten as follows
\beq
\Gamma(t)=\frac{\lambda }{m_0^2}\Lambda(t)\int d^3u\int d^3v\, m(\boldsymbol{u})m(\boldsymbol{v})\left(2e^{-\frac{(\boldsymbol{u}-\boldsymbol{v})^2}{4r_C^2}}-e^{-\frac{(\boldsymbol{u}-\boldsymbol{v}-\boldsymbol{\Delta})^2}{4r_C^2}}-e^{-\frac{(\boldsymbol{u}-\boldsymbol{v}+\boldsymbol{\Delta})^2}{4r_C^2}}\right)\label{gammacont}\,,
\eeq
In the limit of small displacement $\Delta$ in the x direction, the above equation approximates to
\beq
\Gamma(t)=\frac{\lambda}{m_0^2}\frac{\Delta^2}{2r_C^2}\,\Lambda(t)\int d^3u\int d^3v\, \left[1-\frac{(u_x-v_x)^2}{2r_C^2}\right] e^{-\frac{(\boldsymbol{u}-\boldsymbol{v})^2}{4r_C^2}} m(\boldsymbol{u}) m(\boldsymbol{v})\,.
\eeq
If the dimension of the object is much smaller than $r_C$, the terms of the order $(u_x-v_x)/r_C$ can be neglected, and the integrals simply give the total mass squared, thus leading to Eq.~\eqref{n2Napp}.

\section*{Appendix C: Heating effects}
We show that phonon excitations do not contribute in a relevant way to the collapse rate.
It is reasonable to assume that electronic links in the considered minimal setup  together amount  to a copper wire with the following dimensions: length $d=10^{-2}\,{\rm m}$, radius $r=10^{-3}\,{\rm m}$ (volume $V=\pi r^2 d= 3.14\times10^{-8}\,{\rm m}^3$). Such a wire contains   $N_{Cu}=\frac{\mu V}{m}=2.67\times10^{21}$ Cu atoms, where the Cu mass density is $\mu=8.92\times10^3\,{\rm Kg}/{\rm m}^3$, and the Cu atomic weight is $m=1.05\times10^{-25}\,{\rm Kg}$.

We assume that the flowing current excites a phonon with the dimension of the whole wire, the excitation consisting of a rigid displacement of the whole crystal. In order to estimate the displacement induced by the phonon, we first estimate the temperature increase in the wire due to the flowing current. The power dissipated in the wire is $P_{\text{\tiny DISS}}= I^2 R= 1.34\times10^{-5}\,{\rm J}/{\rm s}$, where $I$ is the flash drive current ($I=500$ mA), and the wire resistance is $R=\frac{d\rho}{\pi r^2}=5.35\times10^{-5}\,\Omega$, with $\rho$ the Cu resistivity ($\rho=1.68\times10^{-8}\,\Omega {\rm m}$).
The temperature increase is thus
\begin{equation}
\Delta T=\frac{P_{\text{\tiny DISS}}\, t}{m N C}=1.24\times10^{-8}\,{\rm K}\,,
\end{equation}
where $t=10^{-4}$ s (the longest timescale we have), and $C$ is the Cu thermal capacitance: $C=385\,{\rm K}/{\rm Kg}$.
The phonon displacement $\Delta $ is estimated with the following formula
\begin{equation}
\frac{\Delta }{x_r}=\frac{1}{2}\frac{\Delta T}{T_r}\,
\end{equation}
where $T_r=298$ K, and $x_r=\sqrt{(18h/m\omega_D)(T_r/T_D)}\simeq2\times10^{-11}$~m (where $\omega_D$ and $T_D$ are respectively the Debye frequency and temperature for copper) is the typical atomic displacement in a crystal at room temperature. One thus finds that $\Delta \simeq 4\times 10^{-22}$ m. Recalling that for copper $n= 63.5$, and the total number of displaced copper atoms is $ N_{Cu}\simeq2.67\times10^{21}$, 
substituting  these numbers in Eq.~\eqref{n2N}, in the white noise case one finds that $\Gamma\simeq4.3\times10^{-21}$, thus showing that the heating contribution to the reduction rate is negligible.

\end{document}